# Mn Site Substitution of $La_{0.67}Ca_{0.33}MnO_3$ With Closed Shell Ions: Effect on Magnetic Transition Temperature

L. Seetha Lakshmi, V. Sridharan ∗, D.V. Natarajan, V. Sankara Sastry
and T.S. Radhakrishnan
Materials Science Division, Indira Gandhi Centre for Atomic Research,
Kalpakkam-603 102, India.

**Abstract**:

Mn site is substituted with closed shell ions (Al, Ga. Ti, Zr and a certain combination of Zr and Al) and also with Fe and Ru ions carrying the magnetic moment (S=5/2 and 2 respectively) at a fixed concentration of 5 at. %. Substitution did not change either the crystal symmetry or the oxygen stiochiometry. All substituents were found to suppress both the metal-insulator and ferromagnetic transition temperatures ($T_P(\rho)$ and $T_C$ respectively) to varied extents. Two main contributions identified for the suppression are the lattice disorder arising due to difference in the ionic radii ($|r_{Mn}^{3+}-r_M|$) between the substituent ($r_M$) and the $Mn^{3+}$ ion ($r_{Mn}^{3+}$) and in the case of the substituents carrying a magnetic moment, the type of magnetic coupling between the substituent and that of the neighboring Mn ion.

-------------------------------------------------------------------------------------------------------

∗Corresponding author : V.Sridharan
e-mail : *varadu@igcar.ernet.in*
Abbreviated running title : Mn-site substitution in La-Ca-Mn-O sytem

# 1. Introduction

The hole doped mixed valent $La_{0.67}Ca_{0.33}MnO_3$ compound belonging to the perovskite family exhibits interesting electrical and magneto transport properties especially, the colossal magneto resistance phenomena [1-3]. Though the well known Double Exchange (DE) mechanism could explain the general feature viz., the metallic state coexisting with the ferromagnetic ordering [4-5], subsequent works have shown the need to incorporate polaron (PO) mechanism due to the presence of strong electron phonon coupling to account for the temperature dependence of resistivity in the paramagnetic insulating state [6].

Extensive La-site substitution studies have established the role of hole concentration, average Mn-O bond length ($d_{Mn-O}$) and the average Mn-O-Mn bond angle (<Mn-O-Mn>) in controlling the effective one electron band width and hence the transport properties and the magnetic transition temperatures, $T_C$ [7-10]. Effects of Mn-site substitution with different elements such as Fe, Al, Cr, Co, Ga, Ti etc on the $T_C$ and other physical properties have been reported by many workers [11-20]. It has been observed that all the substituents suppress the $T_C$ although to varied extents. Though the suppression in the $T_C$ has been broadly ascribed to the weakening of the double exchange interaction, the variation in the magnitudes of the suppression rate remains unexplained.

In addition to the average bond length and the bond angle, other factors such as the local structural disorder, the nature of the magnetic coupling between the neighboring spins and the destabilization of the Mn-O network are pertinent to DE and PO mechanisms and have not been addressed by the La-site substitution studies. Effecting appropriate Mn-site substitutions can delineate the different roles and in this paper we address the role of first two effects viz., the local structural disorder and the local magnetic coupling. We effect Mn-site substitutions for a fixed concentration (5 at.%) with a) closed shell ions with smaller (Al, Ti and Ga) as well as larger (Zr) ionic radii compared to that of $Mn^{3+}$ to probe the effects of the local structural disorder arising from the ionic radius mismatch and b) magnetic ions $Fe^{3+}$ (high spin - $t_{2g}^3 e_g^2$) of same ionic radius and $Ru^{4+}$ (high spin state – $t_{2g}^3 e_g^1$), isoelectronic to $Mn^{3+}$ to delineate the effect of magnetic coupling.

In this paper we show that the local structural disorder suppresses the $T_C$. We also show that the substitutions with magnetic ions coupling antiferromagnetically ($Fe^{3+}$) with neighbouring spins suppress the $T_C$. Based on our studies, we argue that enhancement of $T_C$ is possible on substituting a suitable magnetic ion which ferromagntically couple with the Mn ions, although not observed experimentally so far.

**2. Experiment**

The polycrystalline $La_{0.67}Ca_{0.33}MnO_3$ and $La_{0.67}Ca_{0.33}Mn_{0.95}M_{0.05}O_3$ (M=Al, Ga, Ti, Zr, [$Zr_{0.03}Al_{0.02}$], Fe and Ru) were synthesized through standard solid-state reactions. Powder X-ray diffraction in reflection mode was carried out using $Cu_{K\alpha}$ radiation. No impurity phase could be detected from the X-ray diffraction studies. Lattice parameters were obtained by Rietveld refinement. Resistivity measurement in Van der Pauw geometry was carried out [21] using silver paint for contacts. AC susceptibility measurement was carried out using a home built ac susceptometer. The reported Curie transition temperature ($T_C$) corresponds to the onset of ac susceptibility signal, estimated by the tangent method. Oxygen stiochiometry was estimated from weight loss in thermo gravimetric studies under the reduced mixture of Ar (96%) and $H_2$ (4%).

**3. Results and Discussion**

The powder X-ray diffraction patterns of the undoped and the substituted compounds were recorded and representative data are shown in the Figure1. All the patterns except that for [$Zr_{0.03}Al_{0.02}$] could be indexed to orthorhombic phases (space group: P*nma)*. The refined lattice parameters are presented in the Table1. Among the substituents, while Al results in a substantial reduction in the lattice parameters and the unit cell volume, appreciable increases were observed in the case of Zr substitution compared to that of the undoped compound. These observations are consistent with the magnitude and sign of the ionic radius mismatch with respect to $Mn^{3+}$ ion.

Temperature dependence of resistivity and ac susceptibility for undoped and the substituted samples were measured and representative data are shown in the figures 2 and 3 respectively. All the samples showed metal-insulator transitions signified by characteristic peaks in the resistivity at $T = T_P(\rho)$ marking the metal to insulator transition. The values are given in the Table 1 along with the $T_C$, from ac susceptibility measurements. For all the compounds $T_P(\rho)$ matches closely with $T_C$. Below $T_P(\rho)$, there is a sharp change in the resistivity exhibiting a positive $d\rho/dT$. A progressive shift of $T_P(\rho)$ to lower temperatures, an overall increase of resistivity over the entire temperature range and progressive broadening at the peak were observed for the substituents Al, Ti, Ga, Fe, Zr and [$Zr_{0.03}Al_{0.02}$] in that order. Interestingly, the resistivity of Ru substituted compound is lower than that of undoped compound over the entire temperature range of measurement. Similar features, viz., lowering of the $T_C$ and broadening of the transition could also be seen, with the exception of Ru and Zr, in the susceptibility behaviour (Figure 3). In each case, the resistivity in the high temperature paramagnetic region could be fitted to an adiabatic small polaron model [12] and as a representative we show such a fitting for the Al doped compound as an inset in Fig. 2. The thermally activated hopping energy ($E_H$) and the pre-exponential factor (B) are estimated and are given in Table 1. Due to the proximity of the $T_C$ to room temperature, fitting to the small polaron model was not possible for the undoped and Ru doped samples. The $E_H$ for the doped compounds are substantially higher (~140 meV) compared to that of undoped (85 meV) compound. Such an enhancement is in general agreement with earlier observations [12]. However, we could not deduce any systematic.

In earlier works it is reported that $Al^{3+}$, $Ga^{3+}$, $Fe^{3+}$, $Ru^{4+}$, $Ti^{4+}$ [11-13,15,19] and $Zr^{4+}$ [22] substitute the $Mn^{3+}$ site. In figure 4 the $T_C$ suppression rate ($dT_C/dx$) with respect to concentration, for all the substituents is plotted as a function of their ionic radii ($r_M$). Considering the closed shell substituents, viz., Al, Ga Ti and Zr, it is seen that ($dT_C/dx$) varies monotonically with the *absolute magnitude* of ionic radii mismatch ($\Delta r$) between the dopant and the $Mn^{3+}$ ion ($|r_{Mn}^{3+}-r_M|$) barring the Al doped compound. This indicates that the local structural disorder arising from the ionic mismatch weakens the DE interaction, probably through altering the $d_{Mn-O}$ and/or <Mn-O-Mn>, which is reflected in the enhanced values of $E_H$ for the doped compounds. In order to reduce the effect of $\Delta r$, a combination of Zr (larger ionic radius) and Al (smaller ionic radius) with their combined average ionic radius being equal to that of $Mn^{3+}$ was substituted. Though the $T_C$ had marginally increased, substantial suppression was still observed. As of now, the different behaviors of both the Al and [$Zr_{0.03}Al_{0.02}$] substituted systems could not be explained and need further investigation.

Let us consider the $dT_C/dx$ of compounds substituted with magnetic ions namely, $Fe^{3+}$ (S=5/2) and $Ru^{4+}$ (S=2). Going by the ionic mismatch alone, it is expected that Fe substitution should result in the least suppression rate as it has the same ionic radius as that of $Mn^{3+}$. On the contrary, a large suppression rate of 18 K/at.% was observed. In this context, it is also to be noted here that $Ga^{3+}$ having the same ionic size (hence same $\Delta r$) as that of $Ru^{4+}$ results in a suppression rate of 16K/at.%. On the other hand, the suppression rate for the $Ru^{4+}$ substituted system is low, which is 2 K/at.%. Such striking differences in the suppression rates can be rationalised by taking into account the nature of magnetic interaction between the magnetic moments of the substituents and the Mn ion.

It has been shown from magnetisation and Mossbauer measurements that Fe ion couples antiferromagnetically with the neighbouring Mn ions [15], which intrinsically weaken the strength of the DE interaction, and hence decreases the $T_C$. However, $Ru^{4+}$ is isoelectronic to $Mn^{3+}$ and is known to couple ferromagnetically [23]. Such an interaction can substantially compensate for the $T_C$ suppression due to ionic radius mismatch. Thus Ru substitution results in a lower value of $dT_C/dx$ compared to Ga substitution despite the fact that both have same ionic radii. Another interesting feature in the resistivity curve of Ru doped compound is the presence of a hump in the metallic phase (marked by an arrow in the Figure 2) which is much less prominent in the case of the undoped, and the Fe doped compounds. However, no additional feature was observed in ac susceptibility in the temperature range where the hump in the resistivity is seen. High-resolution powder X-Ray diffraction studies on Ru, Fe doped and the undoped system indicated no impurity phase. Hence the low temperature hump in the resistivity is intrinsic to the sample. Such a hump is also reported for Cr doped samples, which is also known to couple ferromagnetically with the neighbouring Mn moments [24]. It appears that the resistivity hump in the metallic region of these systems is a generic feature for ferromagnetically coupled substituents.

**5. Conclusion**

We have investigated the electrical and transport properties of $La_{0.67}Ca_{0.33}Mn_{0.95}M_{0.05}O_3$ (M=Al, Ti, Ru, Ga, Fe, Zr and [$Zr_{0.03}Al_{0.02}$]). Substitutions did neither alter the crystal symmetry nor the oxygen stiochiometry. Both the metal–insulator and ferromagnetic transition temperatures were found to suppress with substitutions to varied extents. Two contributions affecting the transition temperature were identified: 1. Local disorder arising from the ionic radius mismatch between the substituent and Mn ion; the suppression rate found to monotonically increase with the *absolute magnitude* of ionic radius mismatch ($|r_{Mn}^{3+}-r_M|$). 2. Type of magnetic coupling between the magnetic moments of the substituents and Mn ion; ions, which couple antiferromagnetically with Mn ions, suppress the transition temperature.

**FIGURE CAPTIONS**

**Figure 1 :** Powder X- ray diffraction patterns of undoped and $La_{0.67}Ca_{0.33}Mn_{0.95}M_{0.05}O_3$ (M= $Al^{3+}$, $Ru^{4+}$, $Fe^{3+}$ and $Zr^{4+}$).

**Figure 2 :** Resistivity vs temperature of undoped and $La_{0.67}Ca_{0.33}Mn_{0.95}M_{0.05}O_3$ (M= $Al^{3+}$, $Ru^{4+}$, $Fe^{3+}$ and $Zr^{4+}$). Arrow indicates the hump in the resistivity of Ru doped compound. Inset shows the fitting of the resistivity in the paramagnetic region of the Al doped compound to small polaron model. Refer text for details.

**Figure 3 :** ac susceptibility vs temperature of undoped and $La_{0.67}Ca_{0.33}Mn_{0.95}M_{0.05}O_3$ (M= $Al^{3+}$, $Ru^{4+}$, $Fe^{3+}$ and $Zr^{4+}$)

**Figure 4:** Transition Temperature and $dT_c/dx$ vs ionic radius of undoped and $La_{0.67}Ca_{0.33}Mn_{0.95}M_{0.05}O_3$ (M= $Al^{3+}$, $Ti^{4+}$, $Ru^{4+}$, $Ga^{3+}$, $Fe^{3+}$, $Zr^{4+}$ and [Zr:Al]). Pair of arrows indicate ferro (↑↑) and antiferro magnetic (↑↓) coupling between the magnetic moment of the substituent and Mn ion.

**Table:1**

The electronic configuration, the ionic radii of the substituting ions (ref: *R.D. Shannon, Acta. Crystallogra. A 32 (1976) 751)*, refined lattice parameters (numbers in the parenthesis indicates the standard deviation in the last significant digit) and cell volume, Curie transition temperature ($T_C$), the temperature corresponding to the peak in the resistivity ($T_P(\rho)$) marking the metal to insulator transition and the Polaron fit parameters: $E_H$-polaron hopping energy and the pre-exponential factor B for $La_{0.67}Ca_{0.33}Mn_{0.95}M_{0.05}O_3$ (M= $Al^{3+}$, $Ti^{4+}$, $Ru^{3+}$, $Ga^{3+}$, $Fe^{3+}$, $Zr^{4+}$ and [$Zr_{0.03}Al_{0.02}$] ) compounds. ♣ From Literature [12]

| Substituent | Electronic Configuration | Ionic Radius (⊕) | $|r_{Mn}^{3+} - r_M|$ (⊕) | Lattice parameter (⊕) | | | Volume (gm/cm$^3$) | $T_C$ (K) | dTc/dx (K/at.%) | $T_P(\rho)$ (K) | Polaron fit parameters | |
|---|---|---|---|---|---|---|---|---|---|---|---|---|
| | | | | a | b/√2 | c | | | | | $E_H$ (meV) | B (Ω.cm/K) |
| Undoped ($Mn^{3+}$) | [Ar]3d$^4$ | 0.645 | 0.000 | 5.4624(1) | 5.4560(1) | 5.4775(1) | 230.865 | 264.4 | - | 267.32 | ♣84 | ♣2.76x10$^{-6}$ |
| $Al^{3+}$ | [Ne] | 0.530 | 0.115 | 5.4329(5) | 5.4270(7) | 5.4516(5) | 227.318 | 219.77 | 8.9 | 216.10 | 134.24 | 4.97x10$^{-7}$ |
| $Ti^{4+}$ | [Ar] | 0.605 | 0.035 | 5.4635(1) | 5.4582(1) | 5.4740(0) | 230.853 | 134.1 | 26.1 | 143.07 | 142.95 | 5.80 x10$^{-7}$ |
| $Ga^{3+}$ | [Ar] | 0.620 | 0.025 | 5.4512(1) | 5.4452(1) | 5.4666(0) | 229.474 | 200.4 | 13 | 200.30 | 147.63 | 7.11x10$^{-7}$ |
| $Ru^{4+}$ | [Kr]4d$^4$ | 0.620 | 0.025 | 5.4635(3) | 5.4584(4) | 5.4734(3) | 230.840 | 254.02 | 2 | 259.90 | - | - |
| $Fe^{3+}$ | [Ar]3d$^5$ | 0.645 | 0.000 | 5.4552(1) | 5.4636(1) | 5.4542(1) | 229.896 | 177.2 | 17.4 | 182.97 | 145.12 | 8.64x10$^{-7}$ |
| $Zr^{4+}$ | [Kr] | 0.720 | 0.075 | 5.4825(0) | 5.4783(1) | 5.4862(0) | 233.028 | 133.3 | 26.2 | 153.55 | 132.14 | 5.37x10$^{-4}$ |
| [Zr:Al] | - | - | 0.000 | Lattice parameters could not be refined with orthorhombic symmetry (space group:P*nma*) | | | | 149.44 | 23 | 138.19 | 144.82 | 7.70x10$^{-4}$ |

Table 1

L.Seetha Lakshmi *et al*

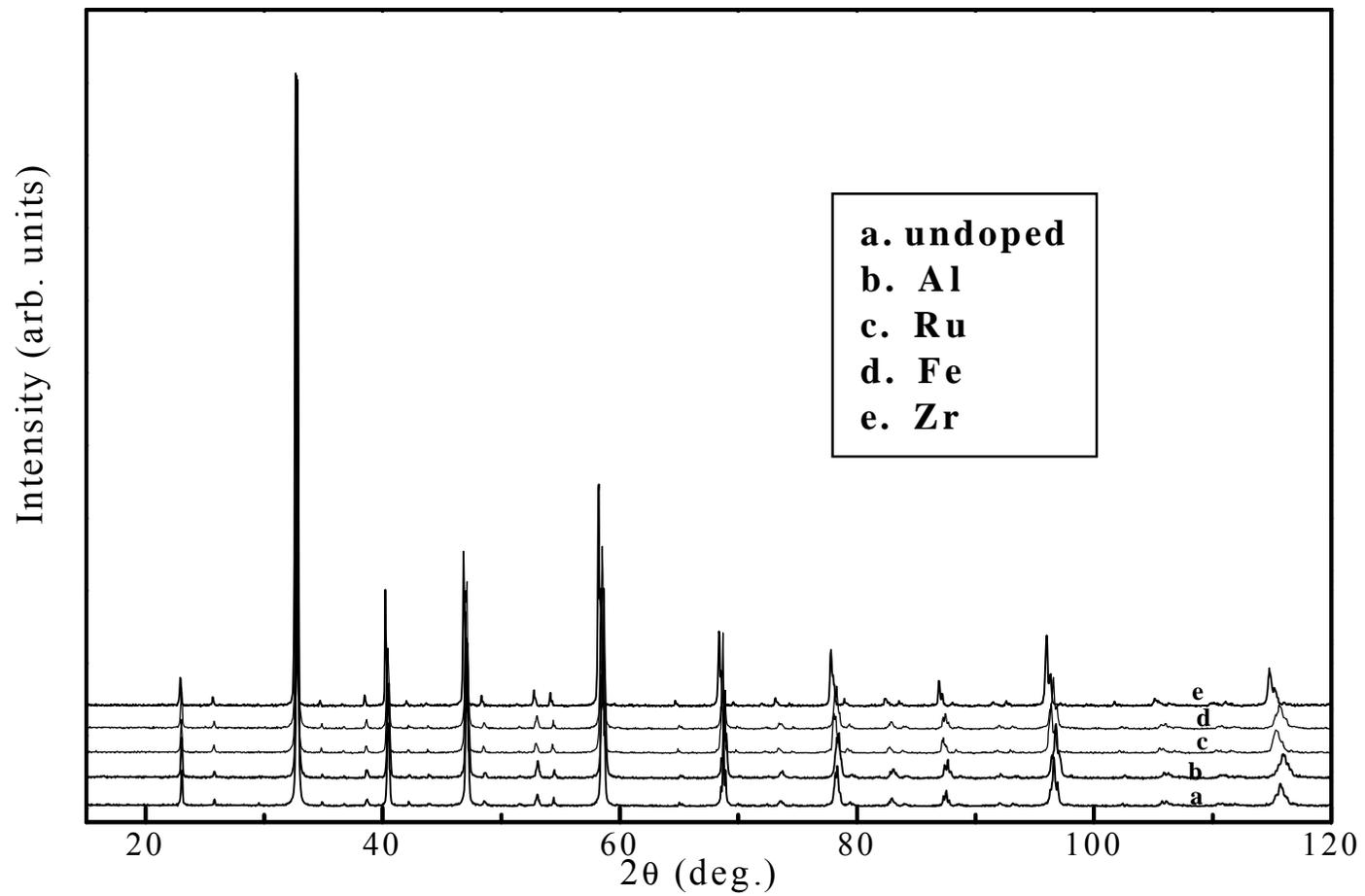

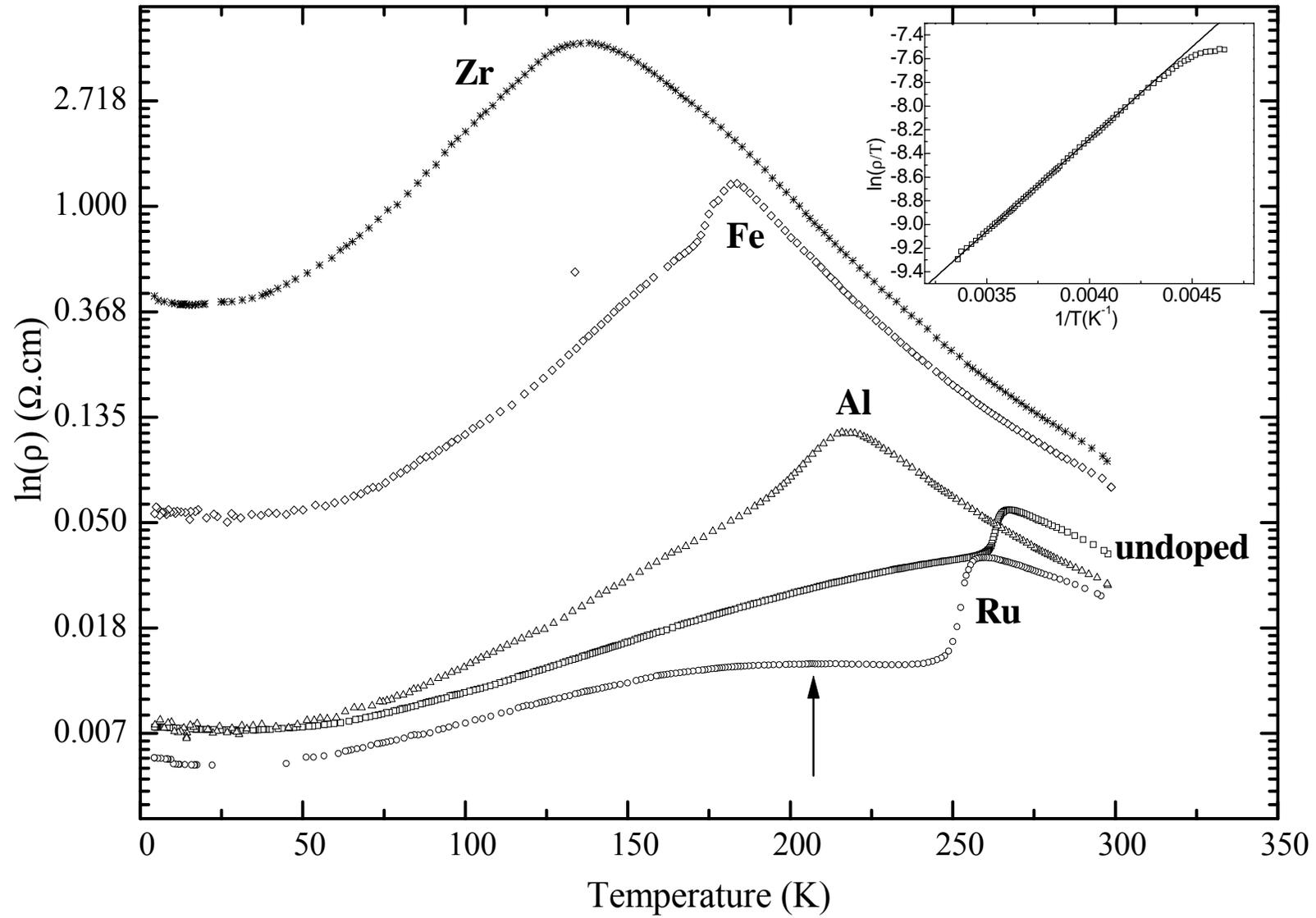

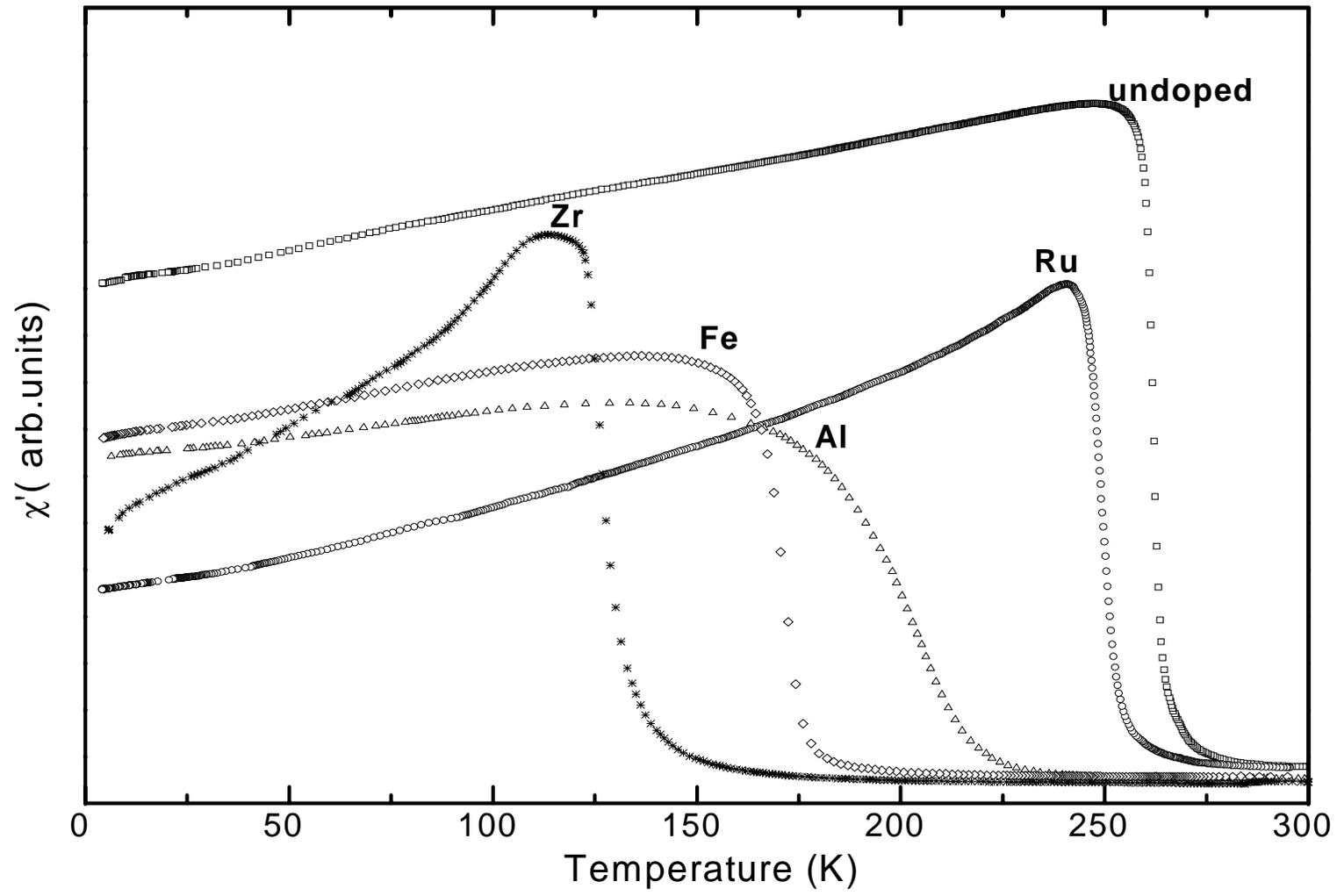

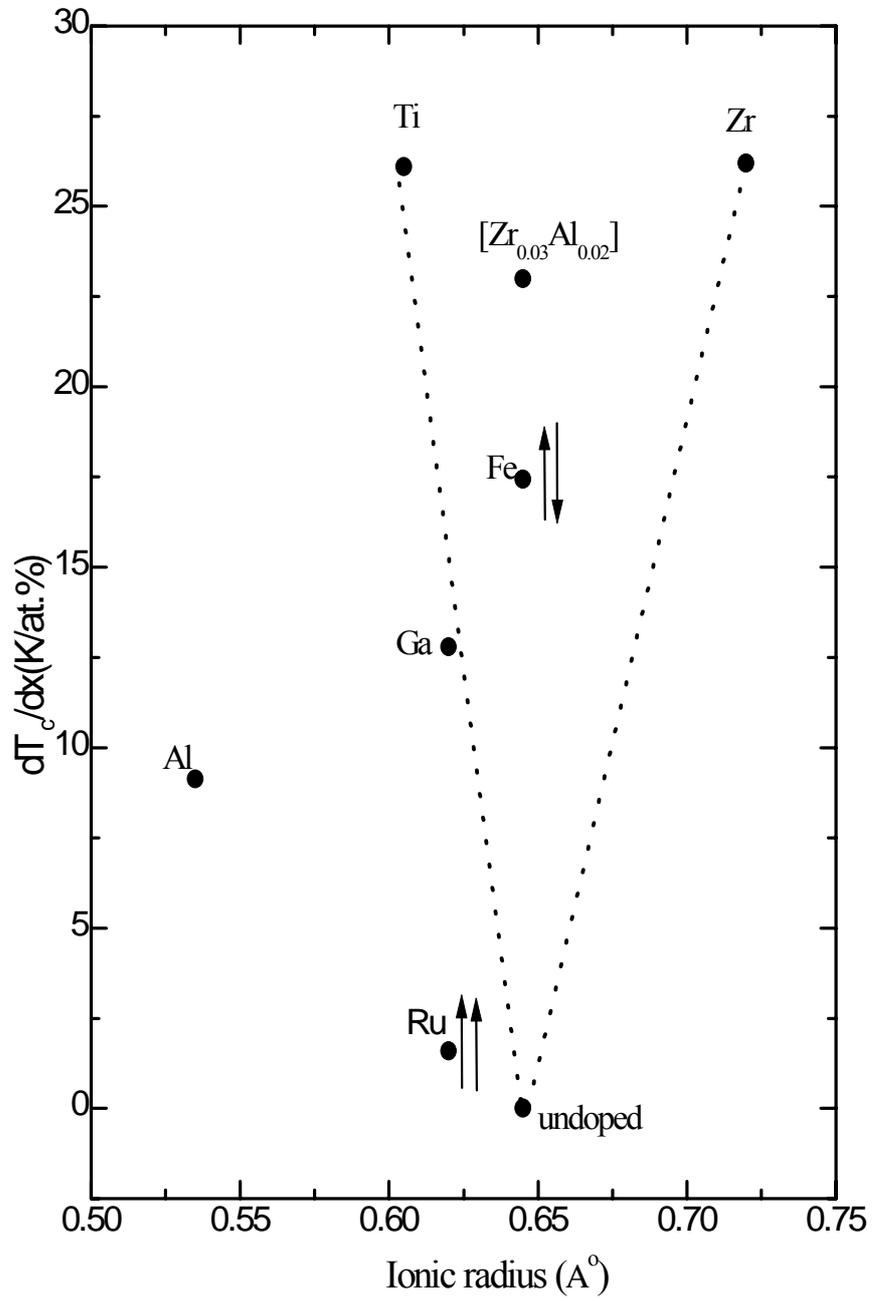